\documentstyle[10pt,aaspp4,Flushrt]{article}

\begin{document}

\def\simg{\mathrel{%
      \rlap{\raise 0.511ex \hbox{$>$}}{\lower 0.511ex \hbox{$\sim$}}}}
\def\siml{\mathrel{%
      \rlap{\raise 0.511ex \hbox{$<$}}{\lower 0.511ex \hbox{$\sim$}}}}
\def\Mesz{M\'esz\'aros~}
\def\dd{{\rm d}} \def\ie{i.e$.$~} \def\eg{e.g$.$~} \def\etal{et al$.$~}
\def\epsel{\varepsilon_{el}} \def\epsmag{\varepsilon_{mag}} \def\eq{eq$.$~}

\begin{center}
\title{MULTI-WAVELENGTH AFTERGLOWS IN GAMMA-RAY BURSTS: \\
           REFRESHED SHOCK AND JET EFFECTS}
\author{A. Panaitescu, P. M\'esz\'aros }
\affil{Department of Astronomy \& Astrophysics,
       Pennsylvania State University, University Park, PA 16802}
\and
\author {M. J. Rees}
\affil{Institute of Astronomy, University of Cambridge, Madingley Road,
       Cambridge, CB3 0HA, U.K.}
\end{center}

\begin{abstract}
We present an analytical approach to calculate the hydrodynamics of
the interaction between a relativistic ejecta and a surrounding medium,
whose evolution serves as a model for Gamma-Ray Burst afterglows.
We investigate the effect of the relevant model parameters on the 
X-ray, optical and radio fluxes, and the effect of a refreshed shock energy 
input and anisotropy in the ejecta on the shape of the light-curves. 
We compare our numerical results to observed afterglows and give a 
quantitative description of the conditions (geometry and physical 
parameters) in the ejecta that are compatible with the light-curves of
the 970508 afterglow, for which a large number of accurate flux
measurements are available.  We find that the radio, optical and X-ray
light-curves of this afterglow can be explained satisfactorily within
the spherically symmetric fireball model, assuming a delayed energy
injection, or by an axially symmetric jet surrounded by a less energetic
outflow. 
\end{abstract}

\keywords{gamma-rays: bursts - methods: numerical - radiation mechanisms:
           non-thermal}

\section{Introduction}

Afterglows from Gamma-Ray Bursts (GRBs) have been observed from a number of
objects at X-ray, optical, and in one case also at radio wavelengths.
Simple analytical models are successful at explaining the major features 
of the light-curves (\Mesz \& Rees 1997, Vietri 1997, Tavani 1997, 
Waxman 1997; Wijers, Rees \& \Mesz 1997). The optical and X-ray light
curves presented by many authors (\eg Pedersen \etal 1998, Piro \etal 
1998, Garcia \etal 1998, Bartolini \etal 1998) have provided evidence for 
occasional departures 
from the basic overall power-law decay behavior. Such departures, as well 
as the possibility of temporal power-law decays that are not 
exclusively determined by the spectral index, have been shown to follow
naturally from fireball models where the radiative regime changes,
the energy is not distributed isotropically in the ejecta (\Mesz, Rees 
\& Wijers 1998), or where the energy input depends 
on the Lorentz factor during the brief injection episode of the central engine, 
leading to refreshed shocks (Rees \& \Mesz 1998). Here we go beyond simple 
analytical asymptotic models, we derive and solve numerically the differential 
equations for the 
dynamics of the afterglow in the general case of a inhomogeneous external medium
and refreshed shock mechanism, and calculate numerically the light-curves
arising in such scenarios. 

Our previous numerical work (Panaitescu \& \Mesz 1997,1998) on simulations of 
light-curves and spectra was based on a hydrodynamic code (Wen, Panaitescu 
\& Laguna 1997) that solves the equations of relativistic hydrodynamics and 
and the shock jump conditions. The energy release mechanisms 
(synchrotron and inverse Compton) were treated as described in Panaitescu \& 
\Mesz (1997). A calculation of the spectra and time history of an afterglow 
from a spherically symmetric shocked fireball is equivalent to computing a quadruple 
integral: over the lab frame time, over the structure of the shocked fluid, 
over the angle relative to the line of sight toward the fireball center (LSC) 
of symmetry and over the electron distribution. The hydrodynamic 
timesteps required to propagate the shell of shocked fluid over times that 
are more than 5 orders of magnitude larger than the shell crossing time and
those necessary for an accurate calculation of the radiative losses lead to 
exceedingly long numerical runs, which are not best suited for an investigation 
of the effects of the large number of model parameters involved in the typical
external shock scenario of GRBs and afterglows (\Mesz \& Rees 1997).  The 
numerical task is even more time-consuming in the case of anisotropic ejecta, 
where a new integral over the azimuthal angle is added.

To acquire computational speed, we have developed a numerical code that 
calculates accurately the evolution of the remnant shell's flow Lorentz 
factor, by solving the equation that gives the evolution of the kinetic energy 
of the remnant during the ejecta--external medium 
interaction, with allowance for an energy injection in the reverse shock 
(Rees \& \Mesz 1998), and radiative and adiabatic losses. Anisotropy of 
the ejecta or of the energy input is included at the simplest level, 
assuming cylindrical symmetry around an axis that is not necessarily the 
same as the LSC. Despite the assumed degree of symmetry in the ejecta, 
the resulting light-curves show a great diversity.  Possible inhomogeneity
of the external medium is considered in the form of a power-law density. 
To simplify 
the energy release treatment, we ignore here the inverse Compton scattering of 
the self-generated synchrotron photons, which is a fairly good assumption, 
substantiated by our previous results (Panaitescu \& \Mesz 1998).
In what follows we describe the analytic treatment of the remnant's dynamics
and energetics, derive analytic light-curves and present our
numerical results. We discuss the effect of model parameters 
on the features of the numerical light-curves, 
and compare them with the afterglow of GRB 970508.

\section {Hydrodynamics of the Remnant and Energy Release}

The most important parameter characterizing the temporal evolution of the 
afterglow is the bulk Lorentz factor $\Gamma$
of the contact discontinuity between the ejecta and 
the swept up external matter. The evolution of $\Gamma$ is 
determined by two main factors, the hydrodynamics of the shell 
(including the energy input, adiabatic losses and the deceleration caused 
by the external medium) and the radiative losses (synchrotron cooling). 

\subsection {Adiabatic remnant}

In the absence of a delayed energy injection and of radiative losses
(adiabatic remnant), at any time, the total energy of the fireball is constant: 
$\dd [M(\Gamma-1)+\Gamma U] = 0$, where $M = M_0 + M_{ex}$ is the total 
remnant mass (the sum of the initial ejecta mass $M_0$ and the swept-up 
mass $M_{ex}$) and $U$ is the co-moving frame internal energy of the 
remnant. The evolution of $U$ is given by the adiabatic losses and by the
heating of the external matter: $\dd U = (\dd U)_{ad} + (\dd U)_{ex}$. 
The jump conditions (Blandford \& McKee 1976) at the forward shock  
imply that $(\dd U)_{ex}=(\Gamma-1)\, \dd M_{ex} c^2$, where $c$ is the
speed of light, therefore the energy conservation can be written as:
\begin{equation}
 (Mc^2+U)\,\dd \Gamma +\Gamma (\dd U)_{ad}+(\Gamma^2-1)\,\dd M_{ex}c^2=0\;,
\label{Gmad}
\end{equation}
where $\dd M_{ex} = 4\,\pi r^2 \rho_{ex}(r)\, \dd r$ and 
$(\dd U)_{ad} =   -(\hat{\gamma}-1)\,U\, \dd (\ln V')_{ad}$. 
The first term in equation (\ref{Gmad}) is the lab-frame change in the 
kinetic energy, the second represents the adiabatic losses and the third 
is the total lab-frame kinetic energy of the shocked external medium, 
including its internal energy. 
In the above equations $r$ is the radial coordinate of the thin remnant,
$\rho_{ex} = \rho_d (r/r_d)^{-\alpha}$, $\rho_d$ is the external medium density
at the deceleration radius $r_d$, defined as the radius where the swept up 
mass is a fraction $\Gamma_0^{-1}$ of the initial fireball mass $M_0$, 
$\Gamma_0$ being its initial Lorentz factor, $\alpha < 3 $ is the index of the 
external matter power-law density, 
$\hat{\gamma}$ is the adiabatic index (maintained close to 4/3 even when
$\Gamma \siml 2$ by the relativistic electrons, provided that the fraction
of electrons that are shock-accelerated is not too much below unity) and $V'$
is the remnant comoving volume. The differential $\dd (\ln V')_{ad}$
refers to that part of the comoving volume that is occupied by the already
shocked fluid, \ie it excludes the change in the comoving volume due to the
sweeping up of the infinitesimal $\dd M_{ex}$.

We assume that the shocked   
external fluid stores most of the internal energy of the remnant and that it
gives most of the afterglow radiation, as the forward shock is more relativistic
than the quasi-newtonian reverse shock, and thus more efficient in converting
kinetic energy into internal and in accelerating high energy electrons that are 
less adiabatic than the electrons accelerated by the reverse shock. 
Therefore, in this work we neglect the dynamical and radiative importance of the 
reverse shock, and leave its treatment for a future, more detailed study.
We also assume that the remnant volume is practically given by the 
volume of the shocked external medium (which is correct if the injected mass is 
not too large compared to the mass of swept up external medium, because the
ejecta compressed between the contact discontinuity and the reverse
shock is denser than the shocked external medium) and that
the comoving density behind the blast wave is equal to that set by the shock jump
conditions. This implies that the comoving density is determined 
solely by $\rho_{ex}(r)$ and $\Gamma$:
$\rho'(r) = (\hat{\gamma}+1)^{-1} (\hat{\gamma}\Gamma + 1)\,\rho_{ex}(r)$. 
Using  $\dd (\ln V')_{ad} = - \dd (\ln \rho')$, the adiabatic losses can be
written as:
\begin{equation}
 (\dd U)_{ad} = -(\hat{\gamma}-1) \left( \frac{\alpha}{r} - \frac{\hat{\gamma}}
 {\hat{\gamma}\Gamma+1} \frac{\dd\Gamma}{\dd r} \right) \, U \, \dd r \;.
\label{dUad}
\end{equation}
The shocked external medium mass is given by
\begin{equation}
 \dd M_{ex} = 3\,\frac{M_0}{\Gamma_0}\,\frac{r^{2-\alpha}}{r_d^{3-\alpha}}\,\dd r\;,
\label{dMem}
\end{equation}
which allows one to calculate the comoving volume $V'= M_{ex}/\rho'$.

\subsection {Delayed Energy Input}

It is possible that the material injected by the cataclysmic event that 
generates the relativistic fireball does not have a unique $\Gamma_0$, 
and that some material is ejected with lower initial bulk Lorentz
factors, down to some limiting $\Gamma_m$. Following Rees \& 
\Mesz (1998), we shall consider that all the ejecta has been released 
impulsively (on a timescale short compared to the afterglow timescale), 
all at the same location, and with a power-law distribution of energy per unit
Lorentz factor $\Gamma_f$: $\dd E_f \propto \Gamma_f^{-s} \dd \Gamma_f$, 
for $\Gamma_m < \Gamma_f < \Gamma_d$, 
where $\Gamma_d < \Gamma_0$ is the Lorentz factor
of the contact discontinuity at $r=r_d$ (for reasons given below, we start
our simulations at $r=r_d$).  The constant of 
proportionality is determined by the total injected energy $E_{inj}$, 
which will be one of the free parameters of the model. The fluid moving 
at lower $\Gamma_f$ lags behind the contact discontinuity  
and catches up with it later, as the fireball is 
progressively decelerated by the interaction with the external fluid. 
From the kinematics of the problem, the Lorentz factor of the ejecta that 
interacts with the reverse shock at radius $r$ is given by
\begin{equation}
 \dd \Gamma_f = -\frac{\Gamma_f}{2}\, \left[\left(\Gamma_f/
                 \Gamma \right)^2 -1 \right] \frac{\dd r}{r} \; . 
\label{Gmin}
\end{equation}
The mass injected in the remnant up to radius $r$ satisfies the 
differential equation
\begin{equation}
 \dd M_f = F_s(\Gamma_f,\Gamma)\, M_{inj}\, \frac{\dd r}{r} \;,
\label{dMin}
\end{equation}
where 
\begin{equation}
 F_s(\Gamma_f,\Gamma) = \frac{s}{2}\, \frac{(\Gamma_f/\Gamma)^2-1}
    {1-(\Gamma_m/\Gamma_d)^s} \left(\frac{\Gamma_m}{\Gamma_f}\right)^s \;.
\label{FSin}
\end{equation}
Here $M_{inj}$ is the total mass that is eventually injected in the remnant,
which corresponds to an $E_{inj}$ given by
\begin{equation}
 E_{inj} = \frac{ s(s-1)^{-1} \left[1-(\Gamma_m/\Gamma_d)^{s-1}\right]
   \Gamma_m+(\Gamma_m/\Gamma_d)^s-1 } {1-(\Gamma_m/\Gamma_d)^s}\, M_{inj}c^2\;.
\label{EINMIN}
\end{equation}
  
As the shocked shell propagates from $r$ to $r+\dd r$, 
the infinitesimal injected mass $\dd M_{f}$ given by equation (\ref{dMin}) 
collides with the shell, increasing the remnant kinetic energy by 
$\dd E_{k,f}$ and its internal energy by $(\dd U)_{f}$. These infinitesimal 
energies can be determined from momentum and energy conservation:
\begin{equation}
 \dd E_{k,f} =\Gamma_{f}[1 -\Gamma^2(1-\beta\beta_{f})] \, \dd M_{f} c^2\;,
\label{dEkin}
\end{equation}
\begin{equation}
(\dd U)_{f} =[\Gamma\Gamma_{f} (1-\beta\beta_{f}) - 1]\,\dd M_{f} c^2 \;,
\label{dUin}
\end{equation}
where $\beta$ represents the velocity in units of $c$.

\subsection {Radiative Losses}

As mentioned before, we consider here that the shocked fluid cools through 
adiabatic expansion and emission of synchrotron radiation from electrons 
accelerated by the forward shock. There could be some contribution to the early 
afterglow light-curve from electrons accelerated by the reverse shock, but this is 
soon overcome by the forward shock emission that shifts toward lower energies, as 
the remnant is decelerated.
We assume that nearly all electrons are shock-accelerated to a power-law
distribution of index $p > 1$, $\dd n'_e \propto \gamma_e^{-p} \dd 
\gamma_e$, for $\gamma_m < \gamma_e < \gamma_M$, where $n'_e$ is the 
co-moving electron number density, $\gamma_e$ is the electron random 
Lorentz factor, and we ignore a possible tail of thermal electrons with 
$\gamma_e < \gamma_m$.
The maximum $\gamma_M$ is determined by the synchrotron
losses during the acceleration timescale, and, for most of the afterglow, 
is several orders of magnitude larger than $\gamma_m$. The minimum 
$\gamma_m$ is set by parameterizing the total energy density stored in 
electrons after acceleration, as a fraction $\epsel$ of the internal 
energy density of the shocked fluid (which is given
by the jump conditions at shock -- Blandford \& McKee 1976), and by the
injection fraction $\zeta$ of electrons that are accelerated at shock. The 
result is
\begin{equation}
 \gamma_m = \frac{p-2}{p-1}\,\frac{1-(\gamma_M/\gamma_m)^{1-p}}
  {1-(\gamma_M/\gamma_m)^{2-p}} \left[1+\frac{\epsel}{\zeta} 
  \frac{m_p}{m_e}(\Gamma-1)\right] \stackrel{p > 2}{\sim} \frac{\epsel}{\zeta}\,
  \frac{p-2}{p-1}\,\frac{m_p}{m_e}\,(\Gamma-1) \;,
\label{gelmin}
\end{equation}
$m_p$ and $m_e$ being the proton and electron masses, respectively.
 
The comoving magnetic field $B$ is assumed to be turbulent and is 
parameterized through the fraction $\epsmag$ of the internal energy that 
is in the form of magnetic field energy,
\begin{equation}
 B = \sqrt{8\pi\, \epsmag \frac{U}{V'}} = B_0 \left[ \epsmag 
     \frac{\hat{\gamma}\Gamma +1}{\hat{\gamma}-1}  \frac{3-\alpha} 
     {3a^3-\alpha a^{\alpha}}\, u \right]^{1/2} \;,
\label{Bmag}
\end{equation}
where $B_0 = (8\pi\,\Gamma_0\rho_d c^2)^{1/2} = 1.94\,(n_0 \Gamma_{0,2})^{1/2}$ G, 
$n_0$ being the external medium particle density at the deceleration radius, in cm$^{-3}$, 
$\Gamma_{0,2}=10^{-2}\,\Gamma_0$ and the non-dimensional variables $a=r/r_d$ 
and $u=U/M_0 c^2$ have been used.

The radiative losses are given by a double integral over the remnant 
volume and the electron distribution,
\begin{equation}
 (\dd U)_{rad} = - \int \dd V'\, \int_{\gamma_m(t')}^{\gamma_M(t')}\,
                \dd n'_e(\gamma_e) P'_{sy}(\gamma_e)\; \dd t' \;,
\label{dUrad}
\end{equation}
which can be calculated for given $B$ and $\dd n'_e(\gamma_e)$ at 
each point in the shocked structure. In equation (\ref{dUrad}), 
$P'_{sy}(\gamma_e) = (4/3)\,\sigma_{Th} c (B^2/8\pi) (\gamma_e^2-1)$ 
is the synchrotron power ($\sigma_{Th}$ being the cross-section for
electron scattering) and $\dd t'= (\Gamma^2-1)^{-1/2}\, c^{-1} \dd r$ gives
the comoving frame time.                                                            
The electron distribution in each infinitesimal ``sub-shell"
within the volume of the shocked fluid is calculated by first initializing it at 
the time $t'$ when the sub-shell is added to the shocked structure and then tracking
the evolution of the electron Lorentz factor $\gamma_e$, subject to adiabatic and 
radiative losses:
\begin{equation}
 \frac{\dd \gamma_e}{\dd a} = -\frac{1}{3}\,\frac{\dd \ln V'}{\dd a}\, (\gamma_e-1)
  -\frac{1}{6\pi} \frac{\sigma_{Th}B^2}{m_e c}\,\frac{\dd t'}{\dd a}\, (\gamma_e^2-1) \;.
\end{equation}

\subsection {Differential Equations and Initial Conditions}

We can now add the contribution of the material injection at the reverse shock 
given by equation (\ref{dEkin}) to the evolution of $\Gamma$ 
(\eq [\ref{Gmad}]), and obtain:
\begin{equation}
 \frac{\dd \Gamma}{\dd a} =
 \frac{ \Gamma_{f}[1-\Gamma^2(1-\beta\beta_{f})] F_s(\Gamma_{f},\Gamma) \omega_M +  
     (\hat{\gamma}-1)\alpha\Gamma u - 3 (\Gamma^2-1) \Gamma_0^{-1} a^{3-\alpha} }
  {a \left[\mu + (\hat{\gamma}^2\Gamma+1)(\hat{\gamma}\Gamma+1)^{-1} u \right]} \;,
\label{dGm}
\end{equation}
where $\mu=M/M_0$ is the dimensionless remnant mass and $\omega_M=M_{inj}/M_0$ is 
the dimensionless total injected mass. We can also obtain the differential 
equation for $U$ by including the internal energy input (\eq [\ref{dUin}]) 
and the radiative losses (\eq [\ref{dUrad}]):
\begin{equation}
 \frac{\dd u}{\dd a}  =
  [\Gamma_{f}\Gamma(1-\beta\beta_{f})-1]F_s(\Gamma_{f},\Gamma) \frac{\omega_M}{a} -
  (\hat{\gamma}-1) \left( \frac{\alpha}{a} - \frac{\hat{\gamma}}{\hat{\gamma}\Gamma+1}
   \frac{\dd \Gamma}{\dd a} \right)\, u + 3 a^{2-\alpha} \frac{\Gamma-1}{\Gamma_0}
   +  \left( \frac{\dd u}{\dd a} \right)_{rad} \;.
\label{du}
\end{equation}
The first term in the numerator of equation (\ref{dGm}) and the first 
term in the right hand side of equation (\ref{du}) are switched off when 
$\Gamma_{f}$, calculated by integrating equation (\ref{Gmin}), drops 
below $\Gamma_m$. The differential equation for the mass of the remnant can 
be obtained with the aid of equations (\ref{dMem}) and (\ref{dMin}):
\begin{equation}
 \frac{\dd \mu}{\dd a} = \frac{3 a^{2-\alpha}}{\Gamma_0} +
                         F_s(\Gamma_{f},\Gamma)\frac{\omega_M}{a} \;.
\label{dmu}
\end{equation}

Solving the hydrodynamics of the remnant is therefore equivalent to
integrating the set of coupled differential equations (\ref{Gmin}), (\ref{dGm}), 
(\ref{du}) and (\ref{dmu}). These equations are valid in any relativistic regime.
If all the released ejecta has the same initial Lorentz factor, then the
fireball is not entirely crossed by the the reverse shock if $r < r_d$ and 
the yet unshocked part of the ejecta and the shocked fluid
move with different Lorentz factors. To avoid unnecessary complications,
we simulate the dynamical evolution of the fireball starting from $r=r_d$ 
and pass over the $r < r_d$ stage by making an appropriate choice of the 
initial conditions at $r=r_d$. These initial conditions are determined by 
the definition of $r_d$.
The energy released before $r_d$, can be safely neglected, so that, 
by equating the sum of the kinetic energy $(1+\Gamma_0^{-1})(\Gamma_d-1)
M_0 c^2$ and the lab frame internal energy $\sim \Gamma_d(\Gamma_d-1)(M_0/\Gamma_0) 
c^2$ at $r_d$ with the initial energy $E_0=(\Gamma_0-1)M_0 c^2$, it is 
straightforward to show that $\Gamma_d = 0.62\,\Gamma_0$ and $U(r_d) = 
(\Gamma_d-1)(M_0/\Gamma_0)c^2 \sim 0.62\, M_0c^2$. Therefore the 
initial conditions are
\begin{equation}
 \Gamma(a=1)=\Gamma_{f}(a=1)=0.62\,\Gamma_0,\quad u(a=1)=0.62,
 \quad {\rm and} \quad \mu(a=1)=1+\Gamma_0^{-1} \;. 
\label{cond}
\end{equation}

\section {Analytical Asymptotic Light-Curves}

The temporal history of the afterglow flux received at Earth can be
calculated analytically by assuming that the ejecta is either 
spherically symmetric or is a jet with axial symmetry, and that $\Gamma$
is power-law in $r$. The last assumption is correct only over a certain
range of times; a different treatment is needed when the remnant 
slows down to non-relativistic speeds. We consider here relativistically 
expanding remnants and, for simplicity, in this section we neglect energy 
injection and restrict our attention to the case when the remnant is 
adiabatic, as it is most 
likely that this stage lasts the longest (Waxman, Kulkarni \& Frail 1998).
Electrons can be either radiative or adiabatic. The former case is compatible
with the assumption of an adiabatic remnant provided that electrons are not
re-energized after shock acceleration or that $\epsel$ is small enough that
most of the internal energy is stored in protons and magnetic fields and lost
adiabatically. In what follows, we denote by $\gamma_m$ the minimum Lorentz
factor of the electrons (in the power-law distribution) that have just been
accelerated, \ie those electrons that are located very close the forward shock.

For a relativistic and adiabatic remnant, to a good approximation, the 
Lorentz factor of the contact discontinuity evolves as (\Mesz \etal 1998)
\begin{equation} 
 \Gamma \sim \Gamma_d (r/r_d)^{-(3-\alpha)/2} \;,
\label{Gm}
\end{equation}
where we consider only the $\alpha < 3$ case. 
The definition of $r_d$ gives $r_d \propto (E_0 n_d^{-1} \Gamma_0^{-2})^{1/3}$,
where $n_d$ is the external medium particle density at $r_d$, and where a multiplying
factor that has a weak dependence on $\alpha$ has been ignored. For definiteness,
we consider that the power-law behavior of the external medium density is manifested
beyond a radius $R_d$, up to which the external density is almost constant, with $R_d$ 
large enough to cover all the possible values of the deceleration radii encountered
in fireballs with reasonable values of the parameters $\Gamma_0$ and $E_0$. 
This approximation is not affecting the remnant evolution, as the afterglow radiation
is emitted at radii much larger than $R_d$. Using the relationship 
between the observer time $T$ and the lab-frame time $t$, $T \propto 
t/\Gamma^2$, the $T$-dependence of the Lorentz factor is found to be
\begin{equation}
 \Gamma \propto \left[ (E_0/n_d)^{(3-\alpha)/2} \Gamma_0^{\alpha}
          \right]^{1/(12-3\alpha)} T^{-(3-\alpha)/(8-2\alpha)} \;.
\label{Gamma}
\end{equation}
Note that $\Gamma$ is independent on $\Gamma_0$ if the external medium is homogeneous.
As seen by the observer, the (transverse) source size scales as 
$\Gamma T$ if the ejecta is spherically symmetric.
The received flux $F_{\nu_p}$ at the peak $\nu_p$ of the 
synchrotron spectrum is $F_{\nu_p} \propto (\Gamma T)^2 \Gamma^3 
I'_{\nu'_p}$ (\Mesz \etal 1998), where $I'_{\nu'_p}$ is the comoving
synchrotron intensity at the comoving peak frequency $\nu'_p$.

{\bf Radiative Electrons.}
If electrons are radiative, then $I'_{\nu'_p} \propto n'_e (P'_{sy}/\nu'_p) t'_{sy} 
\propto n'_e (\gamma_m B)^{-1}$, where $t'_{sy}$ is the comoving synchrotron cooling
timescale and where we substituted $P'_{sy} t'_{sy} \propto \gamma_m$. 
The magnetic field can be calculated using equation (\ref{Bmag}),                 
where the comoving internal energy is determined by assuming that the lab frame 
internal energy of the adiabatic remnant is a (constant) fraction of the initial energy:
$\Gamma u \propto E_0 \Longrightarrow B \propto \epsmag^{1/2}  
(n_d^{3-\alpha} \Gamma_0^{-2\alpha} E_0^{\alpha})^{1/6} r^{-\alpha/2} \Gamma$.
The same result can be obtained using the jump conditions
at the forward shock: $B^2 \propto \epsmag (\dd U/\dd V') \propto 
\epsmag \rho' \Gamma  \propto  \epsmag \rho_{ex} \Gamma^2$.
By using equations (\ref{gelmin}) (with $p > 2$), (\ref{Gamma}) and $r \propto 
\Gamma^2 T$, one can calculate $I'_{\nu'_p}$ and the evolution of the observed peak 
frequency $\nu_p \propto \gamma_m^2 B \Gamma$.
If observations are made at a frequency $\nu < \nu_p$, the observer ``sees" 
the low energy tail of the synchrotron spectrum, which has a slope of 1/3. 
Then $F_{\nu} = (\nu/\nu_p)^{1/3} F_{\nu_p}$, leading to
\begin{equation}
 F_{rad}(\nu < \nu_p) \propto \epsmag^{-2/3} 
        \epsel^{-5/3} \zeta^{8/3} E_0^{1/3} T  \;. 
\label{FlowR}
\end{equation}
Above $\nu_p$, the synchrotron spectrum has a slope $-p/2$, yielding:
\begin{equation}
  F_{rad}(\nu>\nu_p) \propto \epsmag^{(p-2)/4} \epsel^{p-1} \zeta^{2-p}
                    E_0^{(p+2)/4} T^{-(3p-2)/4} \;.       
\label{FhighR}
\end{equation}
Note that $F_{rad}(\nu)$ is independent on the external medium parameters ($\alpha,n_d$)  
and on the fireball initial Lorentz factor and that it depends strongly 
(powers close to or above 1) on $\epsel$ and also on $\zeta$ if $\nu < \nu_p$ 
and on $E_0$ if $\nu > \nu_p$.

{\bf Adiabatic Electrons.}
If the electrons are adiabatic, then $I'_{\nu'_p} \propto n'_e (P'_{sy}/\nu'_p) 
\Delta' \propto n'_e B \Delta'$, where $\Delta'$ is the comoving
remnant thickness. The product $n'_e \Delta'$ can be calculated
using the fact that $4\pi\,(n'_e/\zeta) m_p r^2 \Delta'$ is the external medium 
mass swept up until radius $r$ is reached. Below the spectral peak
\begin{equation}
 F_{ad}(\nu < \nu_p) \propto \epsmag^{1/3} \epsel^{-2/3} \zeta^{5/3} 
     \left[ \Gamma_0^{-4\alpha} E_0^{2(5-\alpha)} n_d^{2(3-\alpha)}
     \right]^{1/(12-3\alpha)}  T^{(2-\alpha)/(4-\alpha)}  \;.
\label{FlowA}
\end{equation}
For observations made above $\nu_p$, the synchrotron spectrum has a  
slope $-(p-1)/2$, therefore:
\begin{equation}
  F_{ad}(\nu > \nu_p) \propto \epsmag^{(p+1)/4} \epsel^{p-1} \zeta^{2-p}
     \left[ \Gamma_0^{-4\alpha} E_0^{3(p+3)-\alpha (3p+7)/4} n_d^{2(3-\alpha)} 
     \right]^{1/(12-3\alpha)}   T^{-[3(p-1)/4]-[\alpha/(8-2\alpha)]} \;.       
\label{FhighA}
\end{equation}
Generally, the light-curve has a strong dependence 
on $E_0$ and $\epsel$, and also on $\zeta$ if $\nu < \nu_p$ and on $\epsmag$ 
if $\nu > \nu_p$. Other dependences are weak to moderate. For $\alpha \simg 2$, 
the light-curve depends strongly on $\Gamma_0$ and $F_{ad}(\nu < \nu_p)$ increases 
with $T$. The larger $\alpha$, the faster $F_{ad}(\nu > \nu_p)$ decreases with $T$.

{\bf Jets.}
If the ejecta is jet-like (Waxman \etal 1998), 
then equations (\ref{FlowR}) -- (\ref{FhighA}) 
give the correct observed flux in the early afterglow, when the observer 
does not see the edge of the jet or the effect of the sideways escape of the
ejecta (Rhoads 1998). For an observer located at an angle 
$\theta_{obs}$ relative to the jet axis and a jet of half-angular opening 
$\theta_{jet}$ such that $\theta_{obs} \ll \theta_{jet}$, the jet edge is 
seen after $\Gamma$ drops below $\theta_{jet}^{-1}$. In this case, the 
source size is $\propto r \theta_{jet} \propto \Gamma^2 T \theta_{jet}$. 
The light-curve of the afterglow from a jet-like remnant is given by:
\begin{equation}
  F_{rad}(\nu < \nu_p) \propto \epsmag^{-2/3} \epsel^{-5/3} \zeta^{8/3}
      \left[ \Gamma_0^{2\alpha} E_0^{7-2\alpha} n_d^{-(3-\alpha)} 
      \right]^{1/(12-3\alpha)}  T^{1/(4-\alpha)} \;,
\label{FjetlowR}
\end{equation}
\begin{equation}
  F_{rad}(\nu > \nu_p) \propto \epsmag^{(p-2)/4} \epsel^{p-1} \zeta^{2-p}
     \left[ \Gamma_0^{2\alpha}E_0^{3(p+3)-\alpha (3p+10)/4} n_d^{-(3-\alpha)} 
     \right]^{1/(12-3\alpha)}  T^{-(3p/4)-[(2-\alpha)/(8-2\alpha)]} \;,
\label{FjethighR}
\end{equation}
if electrons are radiative and by
\begin{equation}
  F_{ad}(\nu < \nu_p) \propto \epsmag^{1/3} \epsel^{-2/3} \zeta^{5/3}
       \left[ \Gamma_0^{-2\alpha} E_0^{13-3\alpha} n_d^{3-\alpha}
       \right]^{1/(12-3\alpha)}    T^{-1/(4-\alpha)}    \;,
\end{equation}
\begin{equation}
  F_{ad}(\nu > \nu_p) \propto \epsmag^{(p+1)/4} \epsel^{p-1} \zeta^{2-p}
          \left[ \Gamma_0^{-2\alpha} E_0^{3(p+4)-\alpha (3p+11)/4} n_d^{3-\alpha}
          \right]^{1/(12-3\alpha)}   T^{-[3(p-1)/4]-[(6-\alpha)/(8-2\alpha)]} \;,
\label{FjethighA}
\end{equation}
if electrons are adiabatic. 

 A comparison of equations (\ref{FlowR})-({\ref{FhighA}) and     
(\ref{FjetlowR})-(\ref{FjethighA}) shows that the light-curve from beamed ejecta 
rises slower and decays faster than that from a spherical fireball. At the onset
of the $\Gamma < \theta_{jet}^{-1}$ phase, the decay of the afterglow 
steepens by $(3-\alpha)/(4-\alpha)$, yielding a break in the light-curve.
This phase lasts until the escape of the ejecta outside the cone
in which it was initially released becomes important.
Rhoads (1998) has shown that in this case the remnant bulk
Lorentz factor decreases exponentially with radius, and that the decay of 
the afterglow light-curve exhibits another break, but remains a power-law in 
the observer time. It can be shown that the time interval from the onset of the 
exponential phase and the beginning of the non-relativistic phase is 
$2.3\,(\theta_{jet}/10^{\rm o})^2$ times shorter than the duration of the 
$\Gamma < \theta_{jet}^{-1}$ phase and that unless $\theta_{jet} \siml 7^{\rm o}$ 
(which yields a very low probability of observing the afterglow) the
sideways escape phase occurs after the remnant becomes non-relativistic.

{\bf Mixed Electron Radiative Regimes.}
Equations (\ref{FlowR})--(\ref{FjethighA}) 
were derived assuming that all the electrons are either radiative or adiabatic. 
The real situation is more complex, as the more energetic tail of the power-law 
distribution of electrons contains electrons that are radiative and 
contribute more to the received flux at some given frequency 
$\nu \gg \nu_p (\gamma_m)$ than the less energetic $\gamma_m$-electrons, 
which become adiabatic early in the afterglow. 
In fact this is the case with most of the numerical X-ray and optical
afterglows shown in the next section.
If the $\gamma_m$-electrons are adiabatic, the flux at a frequency where the emission
is dominated by more energetic and radiative electrons can be derived
using the $I'_{\nu'_p}$ calculated for adiabatic electrons and the fact
that the spectrum has a slope $-(p-1)/2$ for frequencies above $\nu_p$ and below
the peak frequency of the synchrotron emission from electrons that have a radiative
timescale equal to the adiabatic one, and a slope $-p/2$ above this frequency.
Interestingly, the result is the same as given by equations (\ref{FhighR}) and
(\ref{FjethighR}) for radiative $\gamma_m$-electrons, \ie only the constants of 
proportionality are altered. 

We should keep in mind that the above analytical derivations do not take into account
the shape of the equal arrival time surface, \ie the fact that photons that arrive
simultaneously at detector were emitted at different lab-frame times. 
Moreover, we ignored the fact that there are electrons with
Lorentz factors below the $\gamma_m$ of the freshly accelerated electrons.
For these reasons, the equations (\ref{FlowR})--(\ref{FjethighA}) are of
somewhat limited use and, for more accurate results, one must integrate
numerically the afterglow emission.

\section {Numerical Results}

We have introduced so far the following model parameters: (1) dynamical 
parameters $(E_0;n_d,\alpha;\Gamma_0)$, (2) late energy injection parameters 
$(E_{inj},\Gamma_m,s)$ and (3) energy release parameters 
$(\epsmag;\epsel,p,\zeta)$. To these one must add $(\theta_{jet}, \theta_{obs})$ 
if the ejecta is jet-like. In this section we asses the effect of these
parameters, and consider also the situation where $E_0$ and $\Gamma_0$ have 
an anisotropic distribution in the ejecta, which, in the simplest case, 
introduces one more parameter representing the angular scale of such 
anisotropy. We compare our numerical results to the observed  X-ray (2--10 keV), 
optical ($V$ magnitude) and radio (4.9 GHz) afterglows. We will be looking in 
particular for the parameter values that yield X-ray and/or optical light-curves 
similar to GRB 970508, for which a fairly uniform time coverage is available. 

\subsection {Spherically Symmetric Ejecta}

The simplest case is that of spherically symmetric ejecta with a single
impulsive input of energy. Under the
simplifying assumptions of a relativistic and adiabatic remnant, 
the equations (\ref{FlowR}) -- (\ref{FhighA}) predict the asymptotic
radio, optical and X-ray afterglow. For the range of times considered
here, $\nu_p$ is below optical frequencies and only the radio emission shows 
a peak. This peak generally occurs before $\nu_p$ reaches few GHz, and it is
due to the remnant's transition from the relativistic to the non-relativistic 
regime.  For a homogeneous external medium ($\alpha=0$), radiative electrons and 
$p=2.5$, the above-mentioned equations for a relativistic remnant yield for 
$\nu > \nu_p$ (optical and X-ray fluxes)
\begin{equation}
 F_{O,X} \propto \epsmag^{1/8} \epsel^{3/2} \zeta^{-1/2} E_0^{9/8} T^{-11/8} \;,
\label{frad}
\end{equation}
while radio flux $F_R$ is given by equation (\ref{FlowR}).
If electrons are adiabatic, then 
\begin{equation}
 F_{O,X} \propto \epsmag^{7/8} \epsel^{3/2} \zeta^{-1/2} E_0^{11/8} n_d^{1/2} T^{-9/8} \;, 
\end{equation}
\begin{equation}
 F_R \propto \epsmag^{1/3} \epsel^{-2/3} \zeta^{5/3} E_0^{5/6}  n_d^{1/2} T^{1/2} \;.
\label{fadR}                     
\end{equation}

These analytical approximations are consistent, within their range of
validity, with the numerical results shown in Figure 1. 
For all the the afterglows shown in Figure 1, the $\gamma_m$-electrons become 
adiabatic for $T$ between 0.01 and 10 days, while the remnant enters the 
non-relativistic phase at times between 10 and 300 days, when a slow but 
steady steepening of the light-curves can be seen. 
Figure 1 also shows (with symbols) observational data
taken from IAU Circulars, van Paradijs \etal (1997), Sahu \etal (1997),
Frail (1997), Piro \etal (1998) or inferred from the data presented by 
Galama \etal (1997), Bartolini \etal (1998), and Sokolov \etal (1998).
The numerical results are not meant to be fits to the observational data.

If the optical and X-ray electrons are radiative, 
the afterglows arising from fireballs with larger initial 
energy or energy release parameters are brighter, as implied by equation 
(\ref{frad}). Fireballs with harder electron distributions
lead to afterglows that have a shallower temporal
decay (Figure 1[a2]), as predicted by equations (\ref{FhighR}) and (\ref{FhighA}).
If the electron injection fraction $\zeta$ is sufficiently small,             
the radio afterglow can be undetectable (see \eq [\ref{fadR}]).
The peak of the radio light-curve for the $\zeta=10^{-2}$ afterglow shown in 
Figure 1 is $\sim 10\,\mu$Jy. For the same afterglow the synchrotron peak
from $\gamma_m$-electrons remains above the optical range for several days,
leading to an optical afterglow that is flat for the same duration (see Figure 1[a2]).
The non-detection of radio emission from a remnant that yields observable
optical afterglows could also be due to an inhomogeneous external medium:
the peak of the radio emission of the $\alpha=2$ (pre-ejected wind) case
shown in Figure 1 is $\sim 30 \,\mu$Jy.

 The are some important differences between the light-curves arising   
from a fireball running into a homogeneous external medium and into a pre-ejected wind.     
First note that Figure 1(b1) shows that when the electrons emitting at fixed
frequency (here, in X-ray) are radiative, the afterglow is indeed independent
of the external medium parameters $n_d$ and $\alpha$ (if $\alpha \leq 1$), as predicted 
by equation (\ref{FhighR}). The optical and the radio afterglows
depend on $\alpha$ (this is also true for the  X-ray light-curve if $\alpha > 1$), 
indicating that in these cases the electrons that radiate most of the light in 
the corresponding energy bands are adiabatic (eqs$.$ [\ref{FlowA}] and 
[\ref{FhighA}]). 
In a relativistic remnant, the lab-frame synchrotron cooling 
timescale $t_{sy} \propto \Gamma/(\gamma_e B^2)$ for electrons radiating 
at a peak frequency $\nu_p(\gamma_e) \propto \gamma_e^2 B \Gamma > \nu_p (\gamma_m)$ 
equal to a fixed observing frequency $\nu$ is 
$t_{sy} \propto \nu^{-1/2} (\Gamma/B)^{3/2}$, leading to:
\begin{equation}
 t_{sy} \propto \epsmag^{-3/4} \Gamma_0^{\alpha/2} E_0^{-\alpha/4}
                n_d^{-(3-\alpha)/4} \nu^{-1/2} t^{3\alpha/4} \;.
\label{tsy}
\end{equation}
which is constant in time for a homogeneous external medium,
and increases as $t^{3/2}$ for a pre-ejected wind. 
The adiabatic cooling timescale increases as $t$, if the comoving
density tracks the post-shock density.
Therefore the electrons radiating above $\nu_p$ (\ie in optical and X-ray) 
that are radiative, remain so during the entire afterglow if the external medium 
is homogeneous but eventually become adiabatic if the fireball interacts with a 
pre-ejected wind. The radiative regime of the electrons that emit at a given
frequency changes with the index of the external medium, as implied by the
increase of $t_{sy}$ with $\alpha$ (\eq [\ref{tsy}]), and as suggested by the
light-curves shown in Figure 1(b1) and 1(b2): for $\alpha=0$ the X-ray and optical
electrons are radiative, for $\alpha=1$ only the electrons emitting in X-ray
are radiative, while for $\alpha=2$ they are all adiabatic.
Another important difference between the homogeneous and pre-ejected
external media models is manifested by the duration of the relativistic phase.
From equation (\ref{Gamma}) one can calculate the dependence on model
parameters of the time $T_{nr}$ when the remnant becomes non-relativistic 
($\Gamma \siml 2$):
\begin{equation}
 T_{nr} \propto (E_0/n_d)^{1/3} \Gamma_0^{2\alpha/(9-3\alpha)} \;.
\end{equation}
Obviously, $T_{nr}$ is $\Gamma_0$-independent for $\alpha=0$, 
but it depends strongly on the fireball initial Lorentz
factor in the case of a pre-ejected wind: $T_{nr} \propto \Gamma_0^{4/3}$,
implying that in this case the relativistic phase lasts $\simg 100\, \Gamma_{0,2}^{4/3}$
times longer than in the homogeneous external medium case. The optical brightness
of the $\alpha=2$ afterglow is correspondingly weaker, as shown in Figure 1(b2).

We have ignored the effects of low-frequency synchrotron self-absorption
in the radio range, therefore the Figures 1(a3) and 1(b3) give essentially an upper
limit to the optically thin radio flux expected in this case. A simple 
analytical derivation of the absorption frequency is straightforward 
(\Mesz \& Rees 1997),
but it can easily lead to misleading results, since the fireball contains 
electrons with random Lorentz factors that span more than one order of magnitude, 
all emitting and absorbing the synchrotron radiation.
Taking into account only the newly shocked electrons and ignoring a possible 
low-energy tail of the electron distribution below $\gamma_m$, 
it can be shown (Panaitescu \& \Mesz 1998) that the self-absorption frequency 
is $\nu_{ab} \sim 6.4\,(10\,\epsmag)^{1/5}(10\,\epsel)^{-1} n_0^{3/5}
E_{0,52}^{1/5} T^0$ GHz (at redshift $z=1$) 
for a relativistic remnant and adiabatic electrons, where
$E_{0,52} = E_0/(10^{52}\, {\rm ergs})$ and $n_0 = n_d/(1\,{\rm cm}^{-3})$.
This result is valid until the remnant becomes non-relativistic or until the 
shocked material escape sideways, if the remnant is a jet. Therefore
the optical thickness is $\tau = 1.6$ at 4.9 GHz for $\epsel=\epsmag=0.1$, 
$n_0=1$ and $E_{0,52}=1$, indicating that the radio fluxes shown in Figures 
1(a3) and 1(b3) are overestimated by a factor of $\tau (1- e^{-\tau})^{-1} 
\sim 2$. Post-shock mild re-acceleration of the cooling 
electrons or an electron (acceleration) injection fraction $\zeta$ below unity 
can further decrease the radio flux by reducing the number of the 
low energy electrons in the remnant.

In its simplest form considered in Figure 1, the fireball shock model obviously
cannot explain departures from the power-law decay, such as observed 
in the optical afterglow of GRB 970508 near $T \sim 2$ days. A brightening of the 
afterglow may arise if there is a delayed energy input, as illustrated in Figure 2. 
The energy injection index $s$ was set equal to a large value so that the input 
resembles a second relativistic shell that catches up with the initial fireball. 
For a delayed energy input $E_{inj}$ comparable to or larger 
than the energy of the remnant $E_0$, the light-curves exhibit a bump at the 
time of interaction between the two shells. The larger $E_{inj}$ is, the 
more prominent is the resulting bump. For lower $\Gamma_m$, the collision takes 
place later, and this might explain a secondary departure from a 
power-law, apparent in the optical afterglow of GRB 970508 at $T \simg 50$ days. 
(the flattening of the light-curve could also be due to a constant 
contribution of the host galaxy -- Pedersen \etal 1998). 

In Figure 2, the minimum Lorentz factor $\Gamma_m$ was chosen such that the 
numerical light-curve exhibits the brightening observed in the 970508 optical 
afterglow after $T=1$ day. All light-curves shown in Figure 2 were calculated 
using the same fireball initial energy $E_0=6 \times 10^{51}$ ergs, delayed 
energy injection (from refreshed shocks) $E_{inj}=3\,E_0$ (yielding a total 
energy $E_0+E_{inj}=2.4 \times 10^{52}$ ergs), and $\Gamma_m=11$, and the same 
set of parameters $(n_0,\alpha;\epsmag;\epsel)$.  The model shown with dotted 
lines corresponds to constant parameters $p$ and $\zeta$, chosen such that the 
slope of the late optical power-law decay and the early time radio fluxes are 
close to the observed ones. The corresponding X-ray afterglow is too faint,
while the early optical and late radio afterglows are too bright. Generally, 
such discrepancies cannot resolved by adjusting the dynamical parameters
$(E_0,E_{inj},\Gamma_m;n_0,\alpha)$ or the energy release parameters $(\epsmag;
\epsel)$, as changes in these parameters alter the multi-wavelength light-curves 
in a similar fashion. However, a physically plausible possibility is that changes
occur in the parameters $p$ and $\zeta$ which determine the shape of the 
synchrotron spectrum, and these can alter the light-curve in a given band 
without significant changes in other bands. 

For times $T \simg 0.3$ days in Figures 1 and 2, the synchrotron peak $\nu_p$ 
is below the optical band, so that the relative intensity of the optical and 
the X-ray fluxes is determined only by the slope of the spectrum above $\nu_p$. 
This suggests that a brighter X-ray afterglow and a dimmer optical light-curve 
can be obtained by using a flatter electron index $p$, as illustrated by the 
early X-ray and optical fluxes shown with dashed lines in Figure 2. 
If $p$ were held constant at 1.4 during the entire afterglow, the resulting 
optical light-curve would decay much slower than for $p=2.3$ (see \eq 
[\ref{FhighR}]), and thus would be clearly inconsistent with the observational 
data.  A better simultaneous fit of the X-ray and optical afterglows
can be obtained if one assumes that the electron index changes during the
evolution of the remnant. 
In the model shown with dashed lines in Figure 2, we considered that 
the index $p=1.4$ is constant until the second shell of ejecta catches up 
with the fireball ($T \sim 2$ days), and changes to $p=2.3$ at the end of 
the collision between the two shells. 
The indices $p$ before and after the delayed energy input were chosen so that 
the numerical result fits the early X-ray to optical emission ratio and 
the decay of the observed optical light-curve. 
The electrons that radiate most of the $V$-band light shown in Figure 2 are radiative, 
with some smaller contribution from the adiabatic $\gamma_m$-electrons, implying
that the optical spectrum should have a slope close to $-p/2$. The change
from $p=1.4$ to $p=2.3$ at $T \sim 2$ days is consistent with the optical
spectral slopes reported by Djorgovski \etal (1997) ($-0.65 \pm 0.30$ at $T \sim 1$
day), Metzger \etal (1997) ($-0.9 \pm 0.3$ at $T \sim 2$ days), and Sokolov
\etal (1998) ($-1.1$ for $T$ between 2 and 5 days).

The radio afterglow at times shown in Figure 2 ($T > 3$ days) depends on the
late value of index $p$. Unlike the emission at optical and X-ray energies, 
the emission at radio frequencies is due to all the electrons in the remnant, 
whether they are the first accelerated electrons (that have cooled and emit only 
in radio) or the more energetic, recently accelerated electrons (that radiate at
higher frequencies but extend their emission down into radio through the low
energy synchrotron tail of slope 1/3). The later electrons slightly dominate
the radio emission after $T \sim 10$ days, and lead to the large fluxes
shown with dotted line ($\zeta=0.2$) in Figure 2. This contribution to the
radio emission is diminished if the recently accelerated electrons
have a higher post-shock acceleration Lorentz factor, which can be achieved
if the electron injection fraction $\zeta$ is decreased (see \eq [\ref{gelmin}]).
This is shown by the dot-dashed line in Figure 2, where it was assumed that the 
electron acceleration injection fraction drops from $\zeta=0.2$ to $\zeta=0.05$
when the remnant approaches the non-relativistic regime ($\Gamma \sim 3$). 
At the same time the optical afterglow exhibits a brightening due to the fact
that for $\zeta=0.05$ the synchrotron peak $\nu_p$ is closer to the optical range. 
 
\subsection {Axially Symmetric Jets}

Jet-like outflows obviously reduce the energy requirements of fireballs,
which, if extending over $4\pi$ sr, would require a total energy above
$10^{52}$ ergs to produce the optical fluxes observed in the afterglow of
GRB 970508. In Figure 3(a) we show light-curves arising from jet ejecta whose 
properties are isotropic within the opening angle $\theta_{jet}$. 
From these numerical results, we can draw several conclusions: \\
\hspace*{2mm} (1) As expected, the light-curve decay steepens
   when the observer sees the edge of the jet. This is shown by the departure
   of the dotted line (jet, observer located on the jet axis) from the thick 
   solid line (isotropic fireball) around $T = 6$ days. 
   The smaller $\theta_{jet}$, the earlier such a steepening occurs. \\
\hspace*{2mm} (2) Jets seen at angle $\theta_{obs} < \theta_{jet}$ do 
   not exhibit the rise shown by jets with $\theta_{obs} > \theta_{jet}$. \\
\hspace*{2mm} (3) The larger $\theta_{obs}$, the more delayed and dimmer 
   the afterglow peak. For energies $ E_0 \siml 10^{51}$ ergs, the optical
   emission from jets located at $z=1$ that are seen 
   at an angle $\theta_{obs} > 2\,\theta_{jet}$, is unlikely to be detected. \\
The afterglow that fits best the observations is obtained when energy 
injection is included. The thin solid line in Figure 3(a) is for a 
total delayed energy input 4 times larger than the initial energy of the jet, 
leading to a total available energy of $1.9 \times 10^{51}$ ergs. 

In a more realistic al scenario, the explosive event that generates the ejecta 
may lead to an angle-dependent energy distribution, as considered by \Mesz \etal
(1998). Figure 3(b)
(which is not meant as a fit to the afterglow of GRB 970508), 
shows the effect of such an anisotropic distribution for the particular choice 
where the energy per unit solid angle in the jet is an exponential in the polar 
angle $\theta$: $(\dd E_0/\dd \Omega)(\theta) = (\dd E_0/\dd \Omega)_{axis} 
\exp{(-\theta/\theta_E)}$. For $\theta_E > 0$ the angular energy density 
decreases toward the jet edge while for $\theta_E < 0$ it increases. The 
same angular dependence (with the same angular scale $\theta_E$) was assumed 
for $\Gamma_0$. The initial Lorentz factor has no effect on the light-curve if 
the external medium is homogeneous, as shown in the previous section; 
the motivation for this choice was simply an isotropic mass distribution 
in the ejecta.  To maximize the effect of the anisotropy in the ejecta, 
the observer was placed on the jet axis, and a large jet opening was 
chosen in order to separate this effect from the ``edge effect".  In all 
cases, the energy density at $\theta=0^{\rm o}$ was set to $10^{52}/\pi$
ergs/sr, which leads to the following total jet energies: $E_{0,52}=1$ for 
the isotropic distribution ($\theta_E=\infty$), $E_{0,52}=0.2$ for
$\theta_E=\theta_{jet}/3$ and $E_{0,52}=8.6$ for $\theta_E=-\theta_{jet}/3$.
The light-curve decays agree qualitatively with the results of \Mesz
\etal (1998): if $\dd E_0/\dd \Omega > 0$, then more energy is 
emitted from fluid moving at larger angles relative to the LSC, arriving 
later at detector, and yielding shallower decays than in the isotropic case.
Conversely, if $\dd E_0/\dd \Omega < 0$, then most energy is radiated away 
by the fluid moving close to the LSC; this radiation arrives earlier at 
the detector and leads to steeper light-curve decays.

The case where the observer is located off the jet axis is considered in
Figure 3(c). The parameters $(\dd E_0/\dd \theta)_{axis}$ and $\theta_E$ 
were chosen so that the total energy of the jet is the same in all cases.
The conclusion that can be drawn from Figure 3(c) is that, 
for all other parameters fixed, the light-curve seen by an off-axis
observer is determined mainly by the total energy of the jet and not 
by how this energy is distributed. The ironing out of the details of the 
angular energy distribution in an axially symmetric jet is due to the 
differential relativistic beaming of the radiation emitted by fluid 
moving at angles between $\theta_{obs}-\theta_{jet}$ and $\theta_{obs}+
\theta_{jet}$ relative to the LSC. 

Jets with the parameters given for Figure 3(a) and 3(c) can explain the rise 
and decay of the light-curve of GRB 970508 after $T\sim 0.5$ days.
The emission detected in the early part ($T \siml 0.5$ day) of the optical 
afterglow may be due to some ejected material lying outside the main jet.
In Figures 3(a) and 3(c) we show with dot-dashes lines the emission from  
such a large angle outflow, containing $E_0=7.0\times 10^{50}$ ergs, 
ejected isotropically outside of the central jet of opening angle 
$\theta_{jet}=10^{\rm o}$, whose axis of symmetry is offset by 
$\theta_{obs}=14^{\rm o}$ relative to the LSC.  The sum of the light-curves from 
such a two-component ejecta (central jet and large angle outflow) matches well
the features observed in the afterglow of GRB 970508. The X-ray afterglow can
be fitted as before together with the optical, by making an appropriate choice
of the electron index $p$ in the jet and in the large angle outflow.

\section {Conclusions}

Previous models of GRB afterglow light-curves from cosmological fireball shocks
(\eg \Mesz \etal 1998; Rees \& \Mesz 1998; Sari, Piran \& 
Narayan 1998) have used analytical descriptions based on scaling laws 
valid in the asymptotic limits. These require simplifying assumptions and 
involve various undetermined parameters.  The most important analytical 
results on the afterglow light-curve are given in section \S3. 
They should be used 
with care when making comparisons with observed power-law decays, as electrons 
with different random Lorentz factors can be in different radiating regimes.
Generally, those electrons radiating in optical and X-ray are radiative, while 
those radiating at radio frequencies are adiabatic, at least as long as the 
remnant is relativistic. Moreover, 
the analytical light-curves do not take into account the 
shape of the equal-arrival time surface, and assume that there is a one-to-one
correspondence between the lab-frame time of emission and the detector time.

Numerical calculations provide the environment where the assumptions 
made in analytical derivations can be tested and relaxed, and where results 
are expected to be more accurate. In some cases, like that of a fireball 
in a mildly relativistic regime, or like that of jet ejecta seen at an angle
$\theta_{obs} \neq 0^{\rm o}$, it is cumbersome to obtain analytical results.
At the level of numerical calculations, effects arising from the viewing 
geometry (the equal arrival time surface is not the same as the equal 
lab-frame time surface) or from details of the energy release (\eg an 
accurate tracking of the evolution of the electron random Lorentz factor 
$\gamma_e$) can be properly accounted for. 

We have solved the differential equations for the afterglow evolution and 
integrated the remnant emission to calculate light-curves with different model 
parameters.  Energy injection (refreshed shocks), angular anisotropy and jet-like 
structure of the ejecta allow for a variety of possible behaviors of the numerical 
light-curves, even under the assumption of axial symmetry in the remnant. 
More than one scenario could explain a fairly large fraction 
of the optical data of the GRB 970508 afterglow.  A spherically symmetric 
ejecta with energy injection up to a total energy of $2.4 \times 10^{52}$ ergs, 
or a jet of opening $10^{\rm o}$ seen at an angle of $14^{\rm o}$, in which 
energy is injected up to a total of $\sim 2 \times 10^{51}$ ergs, both located 
at redshift $z=1$, seem to fit most the mentioned afterglow.  Such energies 
are quite conservative in a cosmological scenario, and clearly do not require 
any drastic departures from the simple fireball/firejet scenario. Using a variable 
index of the electron power-law distribution, we obtained a simultaneous good
fit of the X-ray and optical afterglow of GRB 970508.
Synchrotron self-absorption, post-shock re-energization of the
electrons or a decrease in the electron acceleration fraction reduce the radio 
fluxes obtained numerically and yield a better
fit of the radio data, but this comparison is much more uncertain and
model-dependent. 

The first generation of fireball shock models of afterglows were characterized 
by great simplicity and have predicted power-law decaying light-curves.
As one would expect, relaxing some of the assumptions that are usually made in 
the simplest versions of these models, such as isotropy of the ejecta 
or constancy of the parameters that quantify the 
energy release, leads to an improved agreement between numerical results 
and observations.
All of the models presented here still contain simplifying assumptions 
(\eg axial symmetry, power-law delayed energy input), 
which were taken as a starting point in investigating the features 
of the numerical light-curves. While the present data do not require it,
relaxing these assumptions could lead to even more diverse afterglow light-curves.
The variety of behavior exemplified by the models we have discussed
highlights the potential importance of afterglow data as diagnostics for the
dynamics and anisotropy of the ejecta, and emphasizes how much more can be
learned when the sample has grown larger.

\acknowledgements {We are grateful to NASA NAG5-2857, NAG 3801 and the Royal
Society for support, and to R.A.M.J. Wijers and D. Reichart for useful comments.}

%\end{document}

\input Psfig

\begin{figure}
\centerline{\psfig{figure=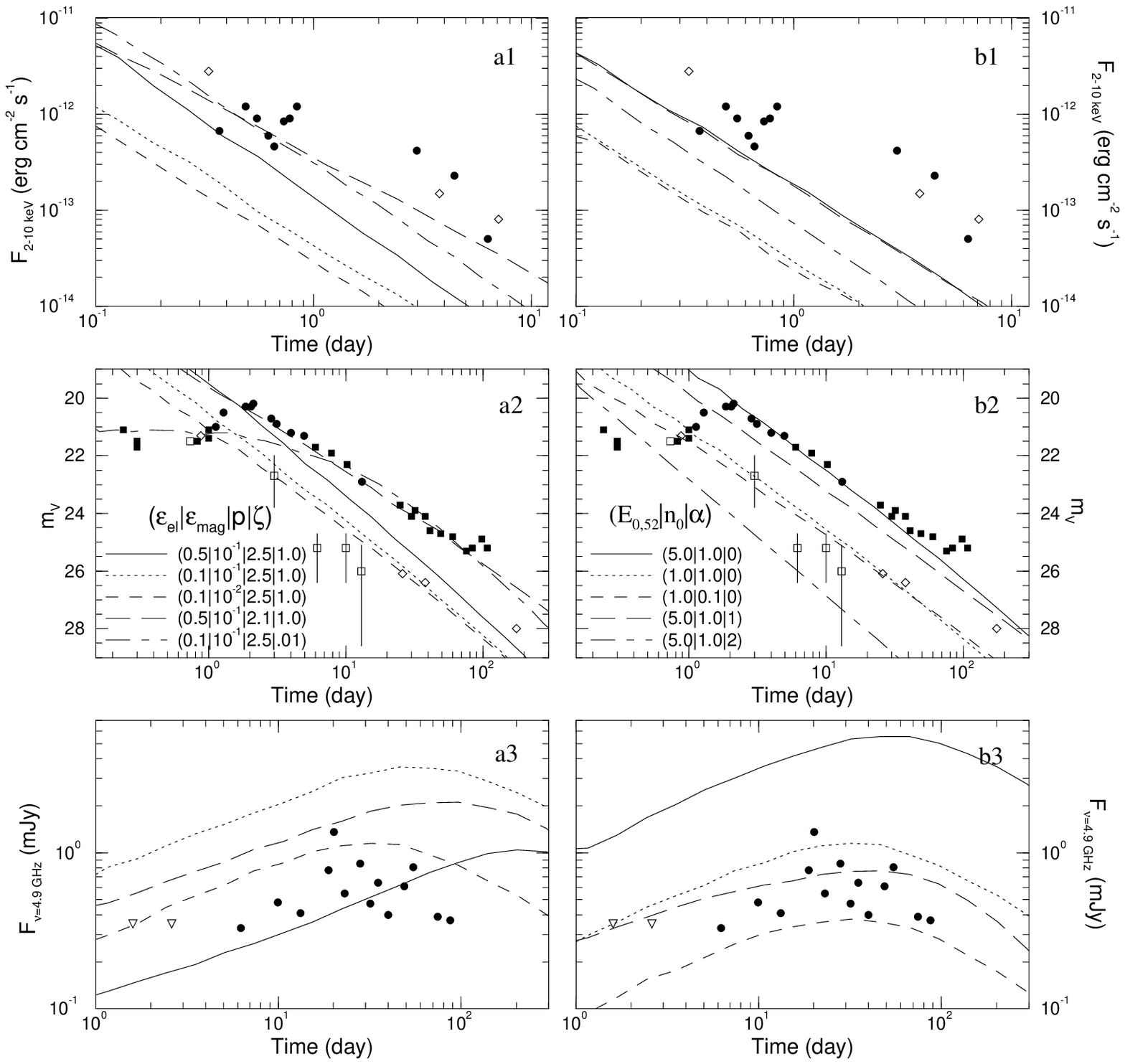}}
\vspace*{-1cm}
\caption {The effect of the energy release (left graphs) and dynamical
(right graphs) parameters on the light-curve from a simple spherically
symmetric fireball, without refreshed shocks. Legends give the parameters 
for each case.  For each curve there is another one that differs in only 
one parameter, allowing assessment of its effect. Other parameters are: 
$E_{0,52}=1$, $n_0 = 1$, $\alpha = 0$ for graphs (a1)--(a3), and 
$\epsmag = 10^{-2}$, $\epsel=0.1$,  $p=2.5$, $\zeta=1$ for graphs (b1)--(b3).
Observational data: open symbols are for GRB 970228, filled symbols for
GRB 970508.  Graphs (a2) and (b2): $V$ magnitudes inferred from $R_C$
magnitudes are shown as squares. Error bars are given only for magnitude 
errors larger than 0.5. Graphs (a3) and (b3):  triangles indicate 
upper limits. The radio light-curves for the $\zeta=10^{-2}$ and $\alpha=2$ 
afterglows have peak fluxes of 10 $\mu$Jy and 30 $\mu$Jy, respectively, 
and do not appear in graphs (a3) and (b3).  A redshift $z=1$ in a $H_0=75\,
{\rm km\,s^{-1} Mpc^{-1}}$, $\Omega=1$ Universe is assumed. The radio fluxes 
plotted are the optically thin upper limits; the inclusion of synchrotron 
self-absorption and/or electron re-energization would lead to lower radio 
fluxes (\S4).} 
\end{figure}
 
\begin{figure}
\vspace*{-4cm}
\centerline{\psfig{figure=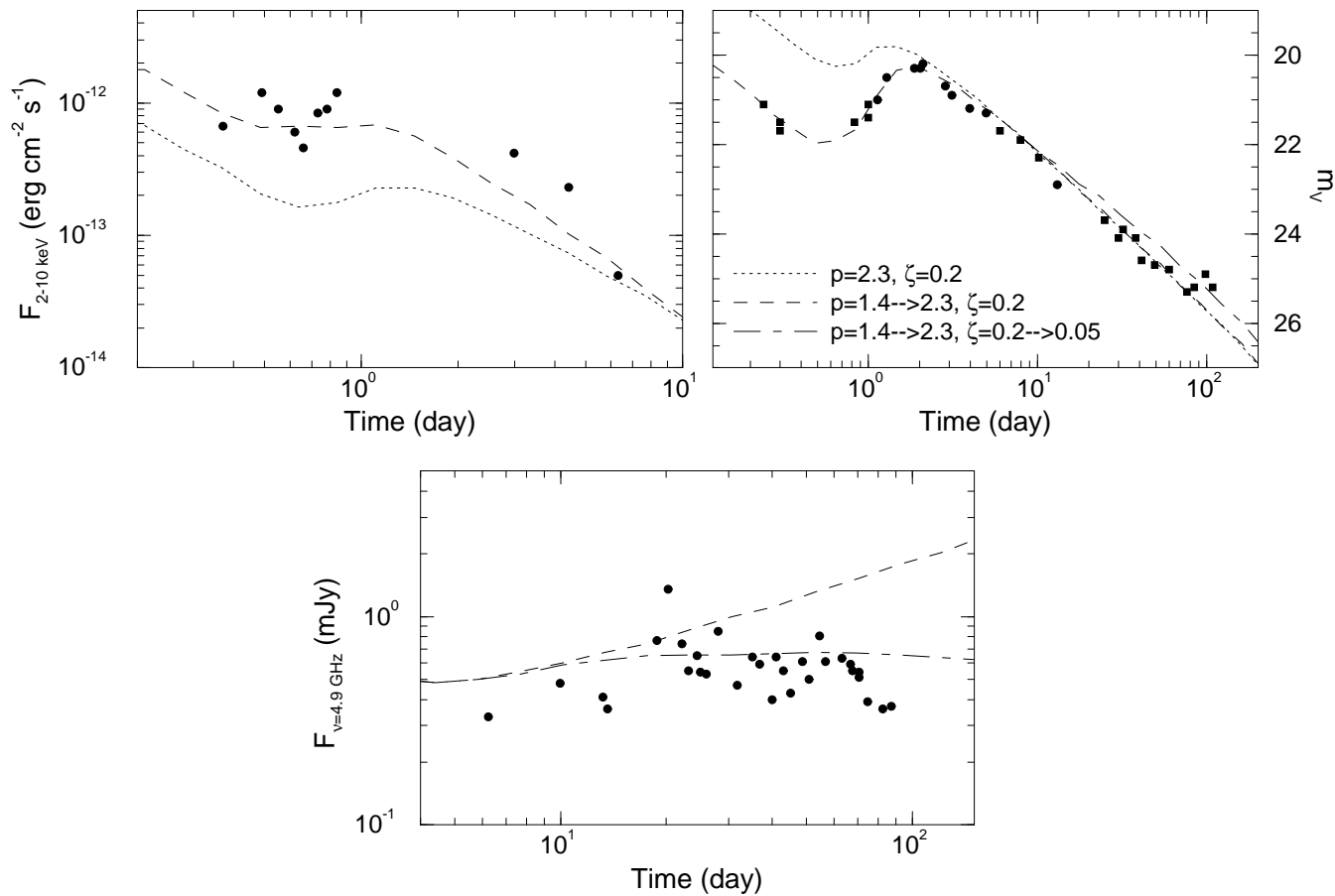}}
\vspace*{-2cm}
\caption {Effect of refreshed shocks in an isotropic fireball, caused by a late 
energy input which is a power law in the Lorentz factor $\Gamma_f$ of the ejecta 
that catches up with the fireball (\S2.2). All models have the same initial and 
injected energies $E_{0,52}=0.6$, $E_{inj} = 3\,E_0$, as well as the same 
minimum Lorentz factor of the delayed energy input $\Gamma_m=11$.
The injection index $s$ has a large value, leading to an impulsive energy
input at $\Gamma_m$ and to a distinctive step-like brightening of the afterglow.
Other parameters are: $\epsmag=0.1$, $\epsel=0.1$, $n_0=1$, $\alpha=0$, $z=1$.
An absorption of $A_V=0.25$ mag (Reichart 1998) at the source redshift was assumed.
The electron index $p$ and acceleration fraction $\zeta$ for each model
are given in the legend of the optical light-curves. They are constant
for the model shown with dotted lines, $p$ changes at the end of the delayed
energy input for the dashed and dot-dashed lines models, while $\zeta$ 
decreases when the remnant ends the relativistic expansion only for the
model shown with a dot-dashed line. 
Symbols represent the data for the GRB 970508 afterglow.}
\end{figure}

\begin{figure}
\vspace*{-2.5cm}
\centerline{\psfig{figure=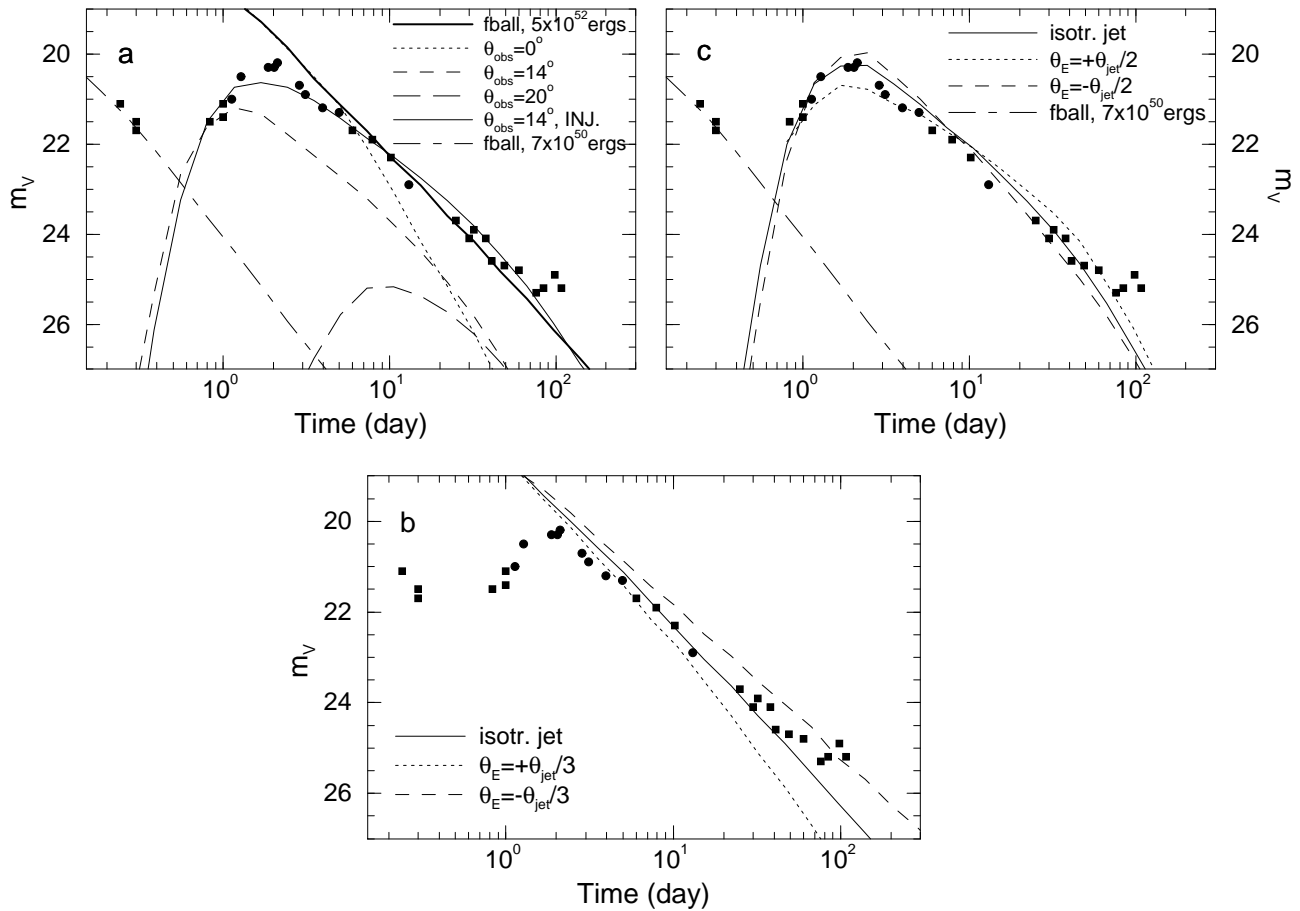}}
\vspace*{-3cm}
\caption{Optical light-curves from jet-like ejecta, compared to data points for 
GRB 970508.
{\bf (a)} An outflow which is isotropic within a jet of opening half-angle 
$\theta_{jet}=10^{\rm o}$, seen at the different angles $\theta_{obs}$, for
$E_0 = 3.8 \times 10^{50}$ ergs, $n_0=1$, $\alpha=0$, $\epsmag=0.1$, $\epsel=0.1$, 
$p=2.5$, $\zeta=1$, $z=1$. For comparison, the afterglow from a spherically 
symmetric remnant with the same parameters, except $E_0 = 5 \times 10^{52}$ ergs 
(yielding the same energy density per solid angle), is also shown (solid thick 
line). A numerical light-curve matching the observational data (solid thin line) 
corresponds to $\theta_{obs}= 14^{\rm o}$ and energy injection characterized by 
$E_{inj}=1.5\times 10^{51}$ ergs, $\Gamma_m=2$ and $s=1.5$.
{\bf (b)} Effect of an anisotropic angular distribution of energy inside a jet 
with $\theta_{jet}=60^{\rm o}$, $\theta_{obs}=0^{\rm o}$, 
$(\dd E_0/\dd \Omega)_{axis}= 10^{52}/\pi$ ergs/sr. Other parameters 
($n_0,\alpha;\epsmag,\epsel,p$) are the same as for graph (a).
The legend gives the angular scale $\theta_E$ (see text).
{\bf (c)} The same jet as in (a) seen at $\theta_{obs}=14^{\rm o}$, but with 
different energy per solid angle distributions. All jets have the same energy 
$E_0 = 1.5 \times 10^{51}$ ergs, isotropically distributed (solid line), 
exponentially decreasing toward the jet edge (dotted line) or exponentially 
increasing toward the edge (dashed line).
Also shown in graphs (a) and (c) with dot-dashed lines is the contribution 
from an ejecta which is isotropic everywhere outside of the jet with opening 
angle $\theta_{jet} = 10^{\rm o}$ and orientation $\theta_{obs}=14^{\rm o}$. 
The isotropic component has an energy $7\times 10^{50}$ ergs (other parameters 
are as for [a]) and can account for the early ($T \siml 1$ day) afterglow 
emission.}
\end{figure}

\end{document}